\documentclass[aps,prl,twocolumn,showpacs,floatfix,amsmath,amssymb]{revtex4}

\usepackage{graphicx}
\usepackage{dcolumn}
\usepackage{bm}

\begin{document}

\preprint{}

\title{Log-normal distribution in growing systems with weighted multiplicative interactions}

\author{Akihiro Fujihara$^{1}$}
 \email{fujihara@yokohama-cu.ac.jp}
\author{Satoshi Tanimoto$^{1}$}
\author{Toshiya Ohtsuki$^{2}$}
\author{Hiroshi Yamamoto$^{2}$}
\affiliation{
 $^{1}$Graduate School of Integrated Science, Yokohama City University, 22-2 Seto, Kanazawa-ku, Yokohama 236-0027, Japan\\
 $^{2}$Field of Natural Sciences, International Graduate College of Arts and Sciences, Yokohama City University, 22-2 Seto, Kanazawa-ku, Yokohama 236-0027, Japan
}

\date{\today}

\begin{abstract}
 Many-body stochastic processes with weighted multiplicative interactions are investigated analytically and numerically. An interaction rate between particles with quantities $x, y$ is controlled by a homogeneous symmetric kernel $K(x, y) \propto x^{w} y^{w}$ with a weight parameter $w$. When $w<0$, a method of moment inequalities is used to derive log-normal type tails in probability distribution functions. The variance of log-normal distributions is expressed in terms of the weight $w$ and interaction parameters. When interactions are weak and a growth rate of systems is small, in particular, the variance is in proportion to the growth rate. This behavior is totally different from that of one-body stochastic processes, where the variance is independent of the growth rate. At $w>0$, Monte Carlo simulations show that the processes end up with a winner-take-all state. 
\end{abstract}



\maketitle

A number of probability distribution functions(PDF) with fat tails of log-normal type have been observed in a wide variety of processes including not only natural but also social and economic phenomena\cite{AB1957, CS1988, LSA2001}. When we conduct a keyword search on "log-normal", we find thousands of literatures in various fields, say, size distributions of small grains\cite{Feltham1957, OC1972, KC1980, KC2004}, aerosols\cite{PG1977, GL1981} clouds\cite{HB1981, Lopez1977, LBRHC1984}, foams\cite{DJC1994}, galaxies\cite{JL2000, Retal2004, HE2004}, the abundance of species\cite{Sugihara1980, Magurran1988, Preston1948}, the number of employees in manufacturing plants\cite{Steindl1965}, the prices of insurance claims\cite{Amoroso1934, Benckert1962}, and the size of farms\cite{Allanson1992}, and so on. In spite of accumulations of a vast amount of data, understandings of log-normal distributions are still quite poor compared with those of power-law distributions. For example, the concepts of self-organized criticality\cite{BTW1988, Jensen1998, Sornette2004} and scale-free network\cite{BA1999, AB2002} are useful in explaining power-law distributions. For log-normal distributions, in contrast, there exist no corresponding explanations. It is well-known that in systems of a single degree of freedom, a multiplicative stochastic process exhibits a log-normal distribution\cite{Kolmogoroff1941, Sornette2004}. In most of actual processes, however, interactions play a significant role and systems must be treated as those with multiple degrees of freedom. In addition, experiments and simulations give only circumstantial evidences. So, there is a real need for analytical derivation of log-normal distributions in many-body systems. To our knowledge, however, the theory of log-normal distributions has not been developed successfully in many-body processes which is irreducible to one-body problem. 

 In recent years, stochastic processes with multiplicative interactions have been attracted considerable attention\cite{BCG2000, EB2002, BK2002, BBLR2003}. Although the processes are simple extensions of the one-body multiplicative stochastic process to many-body processes, it does not exhibit a log-normal distribution but a power law at the tail of PDF\cite{BBLR2003}. In this Letter, we investigate many-body stochastic processes with \textit{weighted} multiplicative interactions analytically. In consequence, we succeed in deriving a log-normal type tail. Furthermore, we find that when interactions are weak and a growth rate of systems is small, the variance of log-normal distribution is in proportion to the growth rate. This behavior is totally different from that of one-body systems, where the variance is independent of the growth rate. 

 We consider a system of $N$ particles with positive quantities $x_i (>0) \; (i = 1, \ldots , N)$. At each time step, the system evolves with a binary interaction between particles labeled by $i$ and $j$ ($i \neq j$), and two quantities $x_i$, $x_j$ are transformed into $x_i'$, $x_j'$ by the rule
\begin{eqnarray}
x_{i}' = \alpha x_{i} + \beta x_{j}, \hspace{5mm} x_{j}' = \beta x_{i} + \alpha x_{j},
\end{eqnarray}
where $\alpha, \beta$ ($>0$) are positive interaction parameters. In the limit $N \to \infty$, the processes are described by the master equation 
\begin{eqnarray}
\frac{\partial f(z,t)}{\partial t} &=& \int_{0}^{\infty}dx \int_{0}^{\infty}dy \ f(x,t) \ f(y,t) \ K(x, y; t) \nonumber \\
				   & & \times \frac{1}{2} \left[\right. \delta( z - ( \alpha x + \beta y ) ) + \delta( z - ( \beta x + \alpha y ) ) \nonumber \\
				   & & - \delta( z - x ) - \delta( z - y ) \left.\right], \label{eq:mastereq}
\end{eqnarray}
where $f(x, t)$ is a PDF of the quantities and $K(x, y)$ is a kernel representing interaction rates. It has been already reported that when a kernel $K$ is constant, in other words, the interaction takes places between randomly chosen particles, the PDF has a power-law tail\cite{BBLR2003}. The exponent of the tail is a continuous function of parameters $\alpha$, $\beta$ and is calculated analytically via a transcendental equation. In many cases, however, the kernel $K$ depends on quantities $x$ and $y$. A typical example is the case where a quantity $x$ represents a mass $M$ or size $R$ of particles. Generally, a transport(diffusion) coefficient and then an interaction(reaction) rate depend on $M$ and $R$. So, it is natural to think what happens in the processes with non-constant kernels. To answer this question, we address the case with a homogeneous symmetric kernel 
\begin{eqnarray}
K(x, y; t) = \frac{x^w \ y^w}{(m_w (t) )^2}, \label{eq:wkernel}
\end{eqnarray}
 where $w$ is a weight parameter and $m_w$ is a $w-$th order moment defined by $m_{w}(t) = \int_{0}^{\infty} x^{w} f(x, t) dx$. Hereafter, we deal with the case $\alpha > 1$ or $\beta > 1$. In this case, the total sum of the quantities increases and the system grows as a whole. When $w$ is positive, interactions between particles with large quantities are accelerated. Thus, the distribution is considered to become wider than the power-law. At $w<0$, the PDF is expected to have a narrower tail. Firstly, the case $w<0$ is investigated by considering an asymptotic moment function of PDF. A log-normal type tail in PDF can be derived analytically and numerical results support this finding. Secondly, Monte Carlo simulations are performed in the case $w>0$. The existence of winner-take-all type distribution is elucidated. 

 As mentioned above, we treat the case where the system grows and does not have a steady state. Then, we attempt to find a scaling solution of Eq.~(\ref{eq:mastereq}) and assume the relations 
\begin{eqnarray}
z = \xi e^{ \gamma t }, \hspace{5mm} f(z, t) = e^{ - \gamma t } \Psi( \xi ), \label{eq:scaling_relation}
\end{eqnarray}
where $\gamma$ is a scaling parameter representing a growth rate of systems. Substituting Eq.~(\ref{eq:scaling_relation}) into Eq.~(\ref{eq:mastereq}), we have a scaled form of the master equation,
\begin{eqnarray}
p \gamma \mu_{p} &=& \int_{0}^{\infty} d\xi_{1} \int_{0}^{\infty} d\xi_{2} \Psi(\xi_{1}) \Psi(\xi_{2}) \frac{(\xi_{1})^{w} (\xi_{2})^{w}}{(\mu_{w})^{2}} \nonumber \\
		 & & \times \left[\right. ( \alpha \xi_{1} + \beta \xi_{2} )^{p} - (\xi_{1})^{p} \left.\right], \hspace{5mm} (p \in \mathbb{R^{+}}), \label{eq:rescaled}
\end{eqnarray}
where $\mu_{p}$ is a scaled $p-$th moment. The parameter $\gamma$ is defined by Eq.~(\ref{eq:rescaled}) at $p=1$, 
\begin{eqnarray}
\gamma \equiv ( \alpha + \beta - 1 )\frac{\mu_{1+w}}{\mu_{1}\mu_{w}}. \label{eq:gamma}
\end{eqnarray}
 Note that Eq.~(\ref{eq:rescaled}) is symmetric with respect to interaction parameters $\alpha$ and $\beta$. Without loss of generality, therefore, we put $\alpha \ge \beta$. To proceed further, we follow the method of moment inequalities which is used to study inelastic hard sphere models by Bobylev \textit{et al.}\cite{BGP2003}. Applying $lemma \ 2$ of \cite{BGP2003} to Eq.~(\ref{eq:rescaled}), we obtain inequalities 
\begin{eqnarray}
\frac{1}{(\mu_{w})^2} \sum_{k=1}^{k_p-1} \left( \begin{array}{c} p \\ k \end{array} \right) \left\{ \left( \frac{\beta}{\alpha} \right)^{p-k} + \left( \frac{\beta}{\alpha} \right)^{k} \right\} \frac{\mu_{k+w} \mu_{p-k+w}}{\mu_{p+w}} \nonumber \\
\le \frac{p \gamma}{\alpha^{p}} \frac{\mu_{p}}{\mu_{p+w}} - \left\{ 1 + \left( \frac{\beta}{\alpha} \right)^p - \left( \frac{1}{\alpha} \right)^{p} \right\} \frac{1}{\mu_{w}} \hspace{22mm} \nonumber \\
\le \frac{1}{(\mu_{w})^2} \sum_{k=1}^{k_p} \left( \begin{array}{c} p \\ k \end{array} \right) \left\{ \left( \frac{\beta}{\alpha} \right)^{p-k} + \left( \frac{\beta}{\alpha} \right)^{k} \right\} \frac{\mu_{k+w} \mu_{p-k+w}}{\mu_{p+w}}, \label{eq:twoinequalities}
\end{eqnarray}
 where $k_p$ denotes the integer part of $(p+1)/2$ for $p>1$. It should be noted that when $p$ is an integer $n \in \mathbb{N}$, inequalities (\ref{eq:twoinequalities}) are reduced to the equality 
\begin{eqnarray}
n ( \alpha + \beta - 1 ) \mu_{n} &=& ( \alpha^{n} + \beta^{n} - 1 )\frac{\mu_{1} \mu_{n+w}}{\mu_{1+w}} \label{eq:moment_relation} \\
				 & & + \sum_{i=1}^{n-1} \left( \begin{array}{c} n \\ i \end{array} \right) \alpha^{i} \beta^{n-i} \frac{\mu_{1} \mu_{i+w} \mu_{n-i+w}}{\mu_{w} \mu_{w+1}}. \nonumber
\end{eqnarray}
 The goal of this analysis is to estimate the asymptotic form of moment function $\mu_{p}$ ($p \to \infty$) when $w<0$. Here, we assume the asymptotic form of $p-$th moment 
\begin{eqnarray}
\mu_{p} = \exp(a p^{2} - b p \ln p + c p), \label{eq:moment_ansatz}
\end{eqnarray}
where $a$, $b$, and $c$ are unknown constants. It is found that Eq.~(\ref{eq:moment_ansatz}) satisfies inequalities 
\begin{eqnarray}
0 \le \left( \begin{array}{c} p \\ k \end{array} \right) \frac{\mu_{k+w} \mu_{p-k+w}}{\mu_{p+w}} \le \tilde{C} \exp\left( - \tilde{A} p + \tilde{B} \ln p \right), 
\end{eqnarray}
for $1 \le k \le k_p $, where $\tilde{A}, \tilde{B}, \tilde{C}(> 0)$ are certain positive constants. It follows that in Eq.~(\ref{eq:twoinequalities}), \textit{l.h.s.} of the first inequality and \textit{r.h.s.} of the second inequality vanish in the limit $p \to \infty$. Hence, the following recurrence equation is fulfilled in the $p-$asymptotic region 
\begin{eqnarray}
p (\alpha + \beta - 1) \mu_{p} = (\alpha^{p} + \beta^{p} - 1) \frac{\mu_{1} \mu_{p+w}}{\mu_{w+1}}, \hspace{1mm} (p \to \infty). \label{eq:asymptotic_moment_relation}
\end{eqnarray}
 Substitution of Eq.~(\ref{eq:moment_ansatz}) into Eq.~(\ref{eq:asymptotic_moment_relation}) leads to 
\begin{eqnarray}
a = \frac{\ln \alpha}{2|w|}, \hspace{5mm} b = \frac{1}{|w|}, \hspace{15mm} \nonumber \\
c = \frac{1}{|w|} \ln \frac{(1 + \delta_{\alpha \beta})\mu_{1}}{(\alpha + \beta - 1) \mu_{w+1}} + \frac{\ln \alpha}{2} + \frac{1}{|w|}, \label{eq:abc}
\end{eqnarray}
 where $\delta_{\alpha \beta}$ is the Kronecker delta. Consequently, the asymptotic solution of moment functions are determined self-consistently. 
\begin{figure}
\resizebox{60mm}{!}{\includegraphics{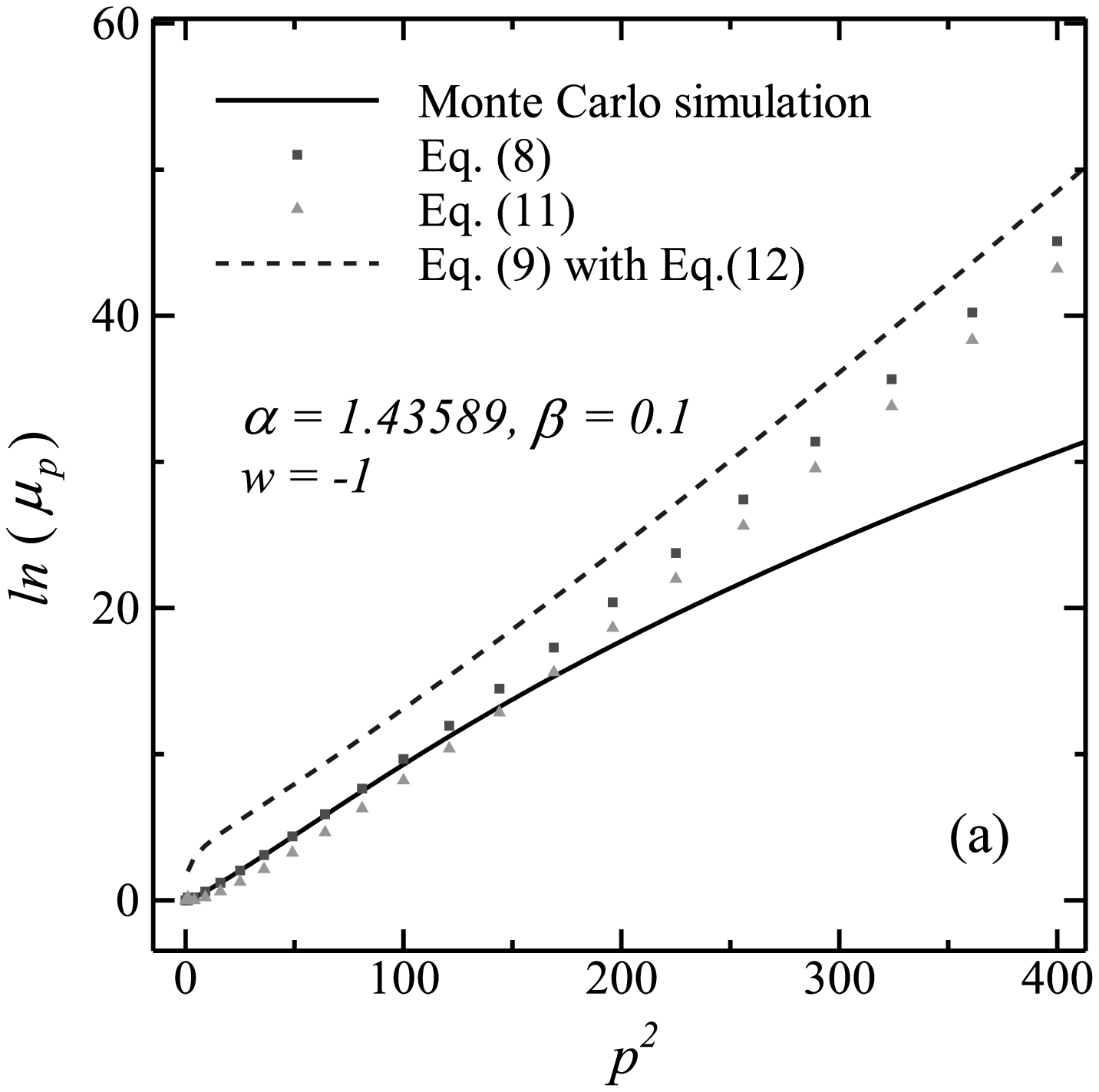}}
\resizebox{60mm}{!}{\includegraphics{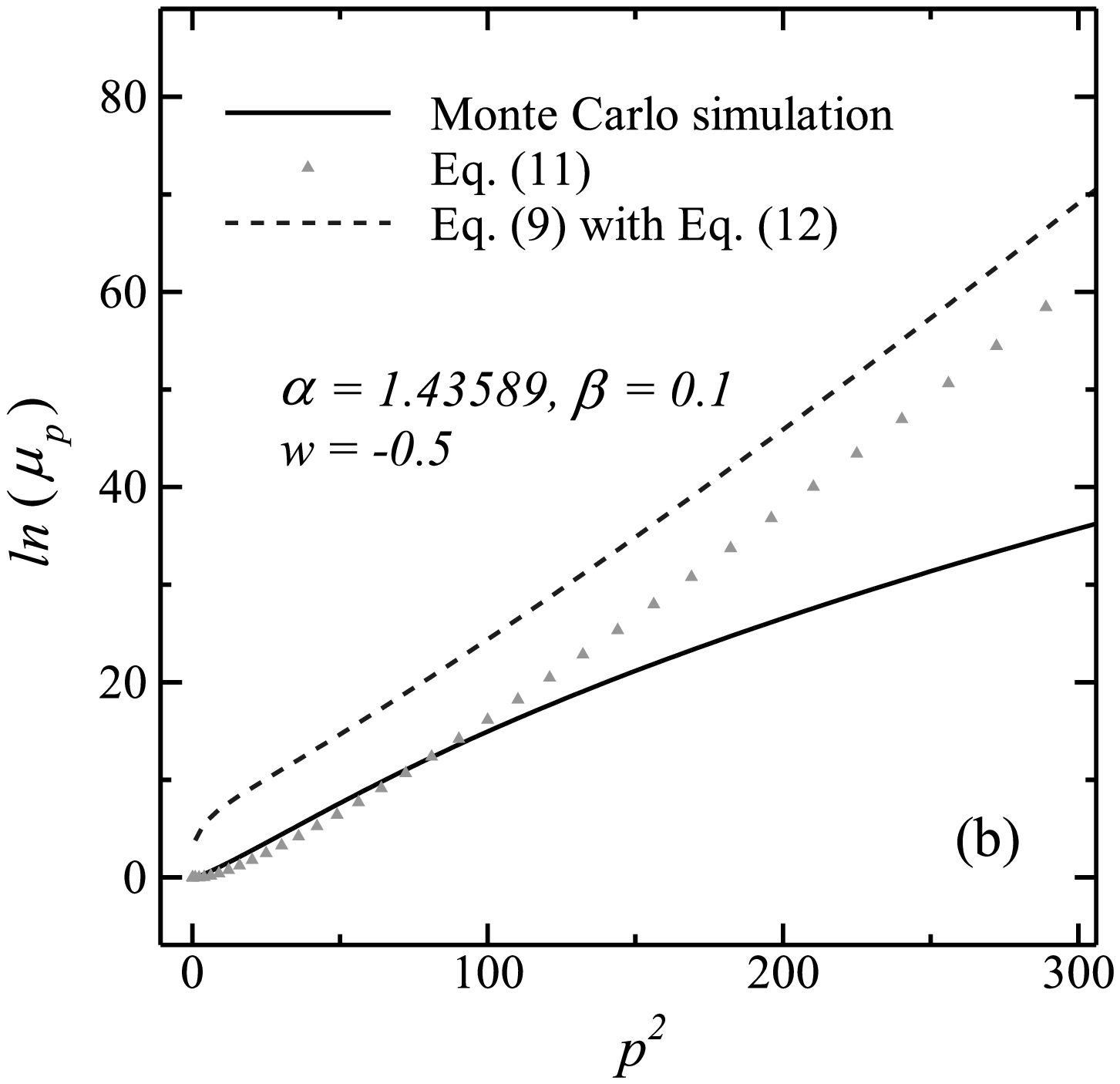}}
\caption{Semi-logarithmic plots of the scaled $p-$th moment $\mu_{p}$ versus $p^{2}$ at a weight parameter $w = -1$ (a) and $-0.5$ (b). Interaction paramaters are $\alpha = 1.43589$ and $\beta = 0.1$. Data of MC simulations are illustrated in solid lines. In MC simulations, the number of particles is $N = 10^{7}$ and the number of interactions is $T = 200 \times N$. Solutions of the recurrence Eq.~(\ref{eq:moment_relation}) and (\ref{eq:asymptotic_moment_relation}) are plotted in square dots ((a) only) and triangular dots, respectively. The results of Eq.~(\ref{eq:moment_ansatz}) with parameters (\ref{eq:abc}) are shown in dashed lines.} 
\label{fig:log_moment_comparison}
\end{figure}
 We compare this result with data of Monte Carlo(MC) simulations and numerical solutions of the recurrence equations (\ref{eq:moment_relation}) and (\ref{eq:asymptotic_moment_relation}). Values obtained by MC simulations are used for the initial values of the recurrence equations. Notice that Eq.~(\ref{eq:moment_relation}) and (\ref{eq:asymptotic_moment_relation}) provide exact results asymptotically. Two typical results at $w=-1$ and $-0.5$ are illustrated in Figs.~\ref{fig:log_moment_comparison}. In the large $p$ region, $\ln(\mu_{p})$ behaves as a quadratic function with a curvature consistent with $a$ in Eq.~(\ref{eq:abc}). MC results agree well at small $p$. Deviation at large $p$ comes from the finiteness of $N$. 
 
 Next, let us consider the shape of PDF. Inverse Mellin transform of the asymptotic moment function Eq.~(\ref{eq:moment_ansatz}) with (\ref{eq:abc}) 
\begin{eqnarray}
\Psi(\xi) = \frac{1}{2\pi i} \int_{l - i \infty}^{l + i \infty} \mu_{s-1} \xi^{-s} ds, 
\end{eqnarray}
gives PDF. It is immediately found that the leading order term of Eq.~(\ref{eq:moment_ansatz}) is transformed into a log-normal distribution. 
\begin{eqnarray}
\Psi(\xi) \simeq \frac{1}{\sqrt{4a\pi} \xi} \exp\left( - \frac{ (\ln \xi)^{2}}{4a} \right), \hspace{5mm} (\xi \gg 1). \label{eq:lognormal} 
\end{eqnarray}
 It is also checked by MC simulations that the PDF is lognormally distributed. It should be emphasized that the variance of the log-normal distribution is determined by $a = \ln (\alpha) / 2|w|$ and is independent of the smaller interaction parameter $\beta$ ($< \alpha$). In many actual phenomena, in addition, interactions are weak and a growth rate is small. In this case($0 < \beta \ll \gamma \ll 1$), Eq.~(\ref{eq:gamma}) and (\ref{eq:abc}) give 
\begin{eqnarray}
a \simeq \frac{1}{2 |w|} \frac{\mu_{1} \mu_{w}}{\mu_{1+w}} \gamma \propto \gamma. \label{eq:agamma}
\end{eqnarray}
 This dependence of the variance $a$ on the growth rate $\gamma$ differs qualitatively from that of one-body processes, where $a$ is independent of $\gamma$. This shows that the log-normal distributions of many-body processes belong to a totally different universality class from those of one-body processes. 
 
\begin{figure}
\resizebox{60mm}{!}{\includegraphics{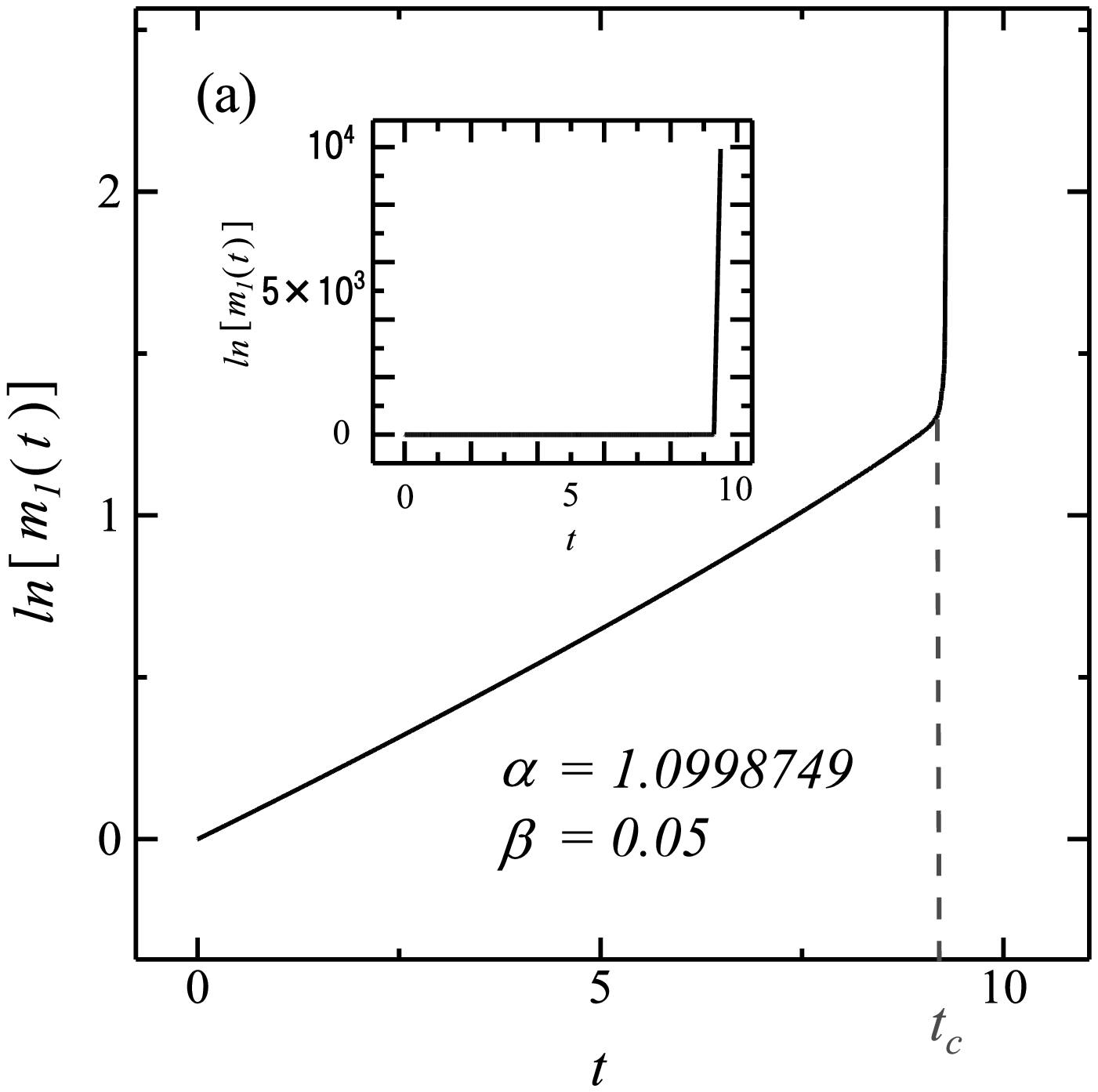}}
\resizebox{60mm}{!}{\includegraphics{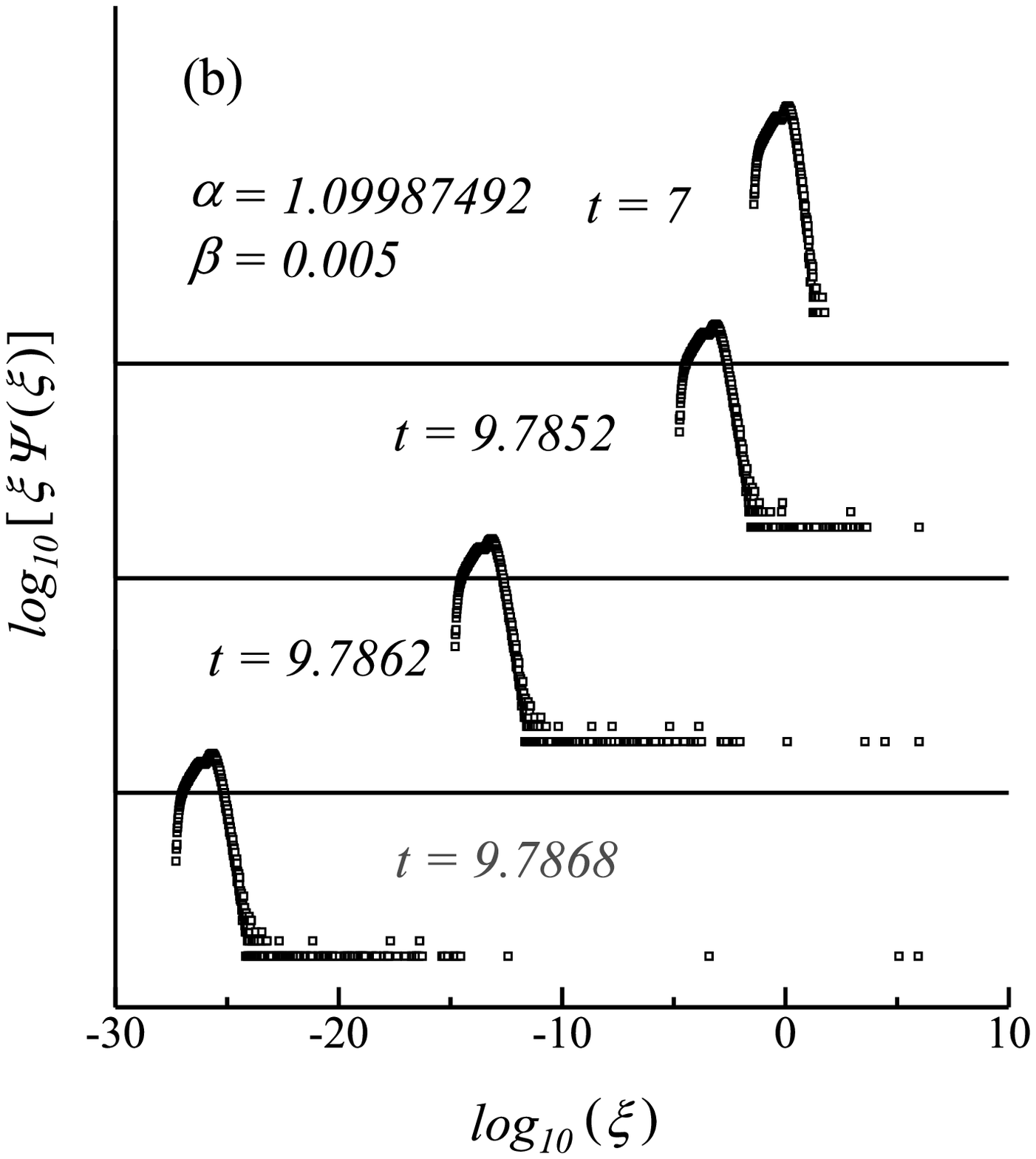}}
\caption{(a) Semi-logarithmic plot of temporal evolution of the first order moment $m_{1}(t)$. The inset shows the divergence of $m_{1}(t)$. (b) Double logarithmic plot of temporal evolution of the PDF. ($\alpha = 1.09987492$, $\beta = 0.005$, and $w = 0.5$.) In MC simulations, the number of particles is $N = 10^{7}$ and unit of time is defined by the number of interaction steps $N/2$.} 
\label{fig:wta}
\end{figure}
 The sign of the weight parameter $w$ is essential to the tail of PDF. When $w=0$, Eq.~(\ref{eq:asymptotic_moment_relation}) reduces to a transcendental equation $(\alpha + \beta - 1) p = \alpha^p + \beta^p -1$, which coincides with the result in \cite{BBLR2003}. When $w>0$, it is suggested by MC simulations that any scaling solution does not hold. This is confirmed by the fact that the growth rate $\gamma$ is not kept constant and $m_{1}(t)$ diverges drastically after a critical time $t_{c}$ as shown in Fig.~\ref{fig:wta}(a). Instead, in the long time limit, the processes end up with a winner-take-all state where two particular particles are selected almost invariably and gain almost all of the total sum of quantities as illustrated in Fig.~\ref{fig:wta}(b). It becomes evident that the power-law distribution($w=0$) emerges at the boundary between the log-normal tail($w<0$) and the winner-take-all state($w>0$). This situation is similar to those of self-organized criticality and scale-free network\cite{BTW1988, Jensen1998, Sornette2004, BA1999, AB2002}, that is, criticality appears at the edge of chaos\cite{Langton1990, Kauffman1993}.

 As far as we know, this work gives the first analytical derivation of log-normal type distribution in many-body processes which is irreducible to one-body problem. It can be concluded that the following two conditions are essential for the emergence of log-normal distributions: (i) systems grow with multiplicative interactions, (ii) interactions between particles with smaller quantities are accelerated. In this Letter, we have treated only the case with a kernel given by Eq.~(\ref{eq:wkernel}). Recently, we have examined the processes with a generic kernel $K(x,y) \propto x^{-a} y^{-b} + x^{-b} y^{-a}$ ($a, b > 0$) and obtained essentially the same results. The details will be reported elsewhere\cite{Fujiharaetal}. 

 The processes are rather simple and expected to become a prototype of certain log-normal type distributions in nature. For example, production methods of fine particles are categorized into two types: breaking-down and building-up. The former may be described by one-body processes as the zeroth approximation. However, the latter must be treated as many-body processes because interactions between particles are essential for the building-up process in general. This suggests that log-normal distributions by building-up methods belong to the universality class of many-body processes with $a \propto \gamma$, while those by breaking-down methods are in the class of one-body processes with $a \sim \gamma^{0} = const$. The difference is expected to be verifiable experimentally.

\end{document}